\documentclass[intlimits,twoside,a4paper]{article}

\usepackage{graphicx}        
\usepackage[cp1251]{inputenc}

\usepackage[eqsecnum]{cmpj3}

\issue{2024}{27}{2}{23703}
\doinumber{10.5488/CMP.27.23703}

\title[Spin-1/2 Ising-Heisenberg distorted diamond chain]%
{Spin-1/2 Ising-Heisenberg distorted diamond chain with
antiferromagnetic Ising and ferromagnetic Heisenberg interactions}
  \author[B. M. Lisnyi]{B. M. Lisnyi\orcid{0009-0002-9061-9458}
  \thanks{E-mail: \email{lisnyj@icmp.lviv.ua}
  }
  }
  \address{Institute for Condensed Matter Physics of the National Academy of Sciences of Ukraine, 1 Svientsitskii St., 79011 Lviv, Ukraine}
\date{Received November 2, 2023, in final form December 15, 2023}

\begin{document}

\maketitle

\begin{abstract}
The exactly solvable spin-1/2 Ising-Heisenberg distorted diamond chain
in the presence of the external magnetic field
is investigated for the case of antiferromagnetic Ising and ferromagnetic \textit{XXZ} Heisenberg interactions.
The influence of quantum fluctuations and the distortion on the ground state,
magnetic and thermal properties of the model are studied in detail.
In particular, it is established that the zero-temperature magnetization curve may involve intermediate
plateaus just at zero and 1/3 of the saturation magnetization.
It is demonstrated that the temperature dependence of the specific heat
reveals up to four distinct peaks at zero magnetic field and up to five distinct peaks at a weak magnetic field.
The physical origin of all observed additional peaks of
the specific heat has been clarified on the grounds of dominating thermal excitations.
We have shown that the quantum fluctuations give rise to an effective geometrical frustration in this chain.
\end{abstract}

\section{Introduction}

Decorated spin chains, which can be exactly solved using the transfer-matrix method
\cite{s5,fis59,roj09,str10,bell13,bax82,aps15},
are of practical importance for the qualitative interpretation and the quantitative description
of magnetic and thermal properties of real solid-state materials.
In particular, several exactly solved Ising-Heisenberg decorated spin chains provide an in-depth understanding of a striking
interplay between geometric spin frustration and quantum fluctuations, which may manifest itself through various
intriguing phenomena such as the appearance of an intermediate plateaux in low-temperature magnetization curves and
the formation of additional maxima in the temperature dependence of the specific heat.
Despite a certain oversimplification, some exactly solved Ising-Heisenberg spin chains afford a plausible quantitative
description of the magnetic behavior of real spin-chain materials
\cite{s0,s3,exp10,sah12,str12,han13,oha14,verkh13,strec20,galis22}.

The natural mineral azurite Cu$_3$(CO$_3$)$_2$(OH)$_2$ provides an experimental realization
of the geometrically frustrated spin-1/2 diamond Heisenberg chain \cite{jes11,hon11}
with spectacular magnetic properties~\cite{kik04,ki05l,ki05ptp,rul08}.
Owing to this fact, a lot of attention has been paid to a research of different versions
of the geometrically frustrated spin-1/2 Ising-Heisenberg diamond chain
\cite{can06,can09,roj11,lis3,ana12,roj12,ana13,gal13,gal14,anan14,pssb14,ohanyan15}.
Although a correct description of magnetic properties of the azurite would
require a modelling based on a more complex spin-1/2 Heisenberg model~\cite{jes11,hon11},
the simplified but still exactly solvable spin-1/2 Ising-Heisenberg diamond chain~\cite{can06,lis3,pssb14}
qualitatively reproduces the most prominent experimental features reported for the azurite such as
an intermediate one-third magnetization plateau as well as the double-peak temperature dependencies of specific heat
\cite{kik04,ki05l,ki05ptp,rul08}.
On the other hand, much less attention has been paid to non-geometrically frustrated cases of
the spin-1/2 Ising-Heisenberg diamond chain.

Exactly solvable Ising-Heisenberg systems can be employed to formulate a theoretical framework for describing
quantum Heisenberg systems. This involves deriving effective Hamiltonians for these systems through
many-body perturbation theory \cite{derzh15,verkh16,verkh21,verkh22}.
In particular, the geometrically frustrated spin-1/2 Ising-Heisenberg diamond chain
was generalized to a spin-1/2 Heisenberg diamond chain by adding a perturbation,
for which a much simpler effective Hamiltonian was obtained within the many-body perturbation theory
to reproduce its low-temperature behavior \cite{derzh15}.
In addition, a quantum spin-1/2 antiferromagnetic Heisenberg trimerized chain was studied
using the many-body perturbation expansion,
which is developed from the exactly solved spin-1/2 antiferromagnetic Ising-Heisenberg diamond chain~\cite{verkh21}.

The main purpose of this work is to examine the ground state and basic thermodynamic properties of
the spin-1/2 Ising-Heisenberg distorted diamond chain
with antiferromagnetic Ising and ferromagnetic Heisenberg interactions.
This chain reduces to a usual spin-1/2 Ising-Heisenberg diamond chain when the asymmetry
of the Ising couplings along the diamond sides vanishes,
while the spin-1/2 Ising-Heisenberg doubly decorated chain is recovered under the extreme case of asymmetry.
The ground state, magnetization process and basic thermodynamic characteristics (specific heat, susceptibility)
of the spin-1/2 Ising-Heisenberg distorted diamond chain will be exactly calculated within the framework
of the transfer-matrix method.
We explore how the quantum fluctuations and
the coupling asymmetry will affect the overall magnetic and thermal behavior.
We show that quantum fluctuations create an effective geometric frustration in this chain,
which causes similar features in the properties of the ground state and basic thermodynamic characteristics
as in the case of the corresponding antiferromagnetic chain with geometric frustration \cite{lis3},
while the features of the temperature dependence of the heat capacity are more intriguing.

\section{Model and its exact solution}

\begin{figure}[!t]
\begin{center}
\includegraphics[width=8cm]{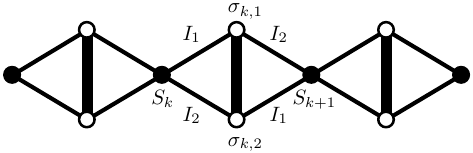}
\end{center}
\caption{A fragment from the spin-1/2 Ising-Heisenberg distorted diamond chain.
The nodal ($S_{k}$, $S_{k+1}$) Ising spins and interstitial ($\sigma_{k,1}$, $\sigma_{k,2}$) Heisenberg spins
belonging to the $k$th primitive cell are shown.
}
\label{fig1}
\end{figure}

Let us begin by considering the spin-1/2 Ising-Heisenberg distorted diamond chain in
the presence of an external magnetic field.
The magnetic structure of the investigated model system is schematically illustrated in figure~\ref{fig1}.
As one can see, the primitive cell in the shape of diamond spin cluster involves two nodal Ising spins
$S_{k}$ and $S_{k+1}$ along with two interstitial Heisenberg spins $\sigma_{k,1}$ and $\sigma_{k,2}$.
The total Hamiltonian for the spin-1/2 Ising-Heisenberg distorted diamond chain in the presence of an external
magnetic field, which contains $N$ primitive cells, reads
\begin{eqnarray}
\hat {\cal H} &{=}& \sum_{k=1}^{N}
\left[S_{k} \left(I_1 \hat\sigma^z_{k,1} + I_2\hat\sigma^z_{k,2} \right) +
S_{k+1} \left(I_2\hat\sigma^z_{k,1} + I_1 \hat\sigma^z_{k,2} \right) \right]
\nonumber\\
&+& \sum_{k=1}^{N} \left( J_1 \hat\sigma^x_{k,1} \hat\sigma^x_{k,2} + J_2 \hat\sigma^y_{k,1} \hat\sigma^y_{k,2}
+ J_3 \hat\sigma^z_{k,1} \hat\sigma^z_{k,2} \right)
- \sum_{k=1}^{N} h \left( S_{k} + \hat\sigma^z_{k,1} + \hat\sigma^z_{k,2} \right),
\label{htot}
\end{eqnarray}
which involves the nodal Ising spins $S_k = \pm 1/2$ and the interstitial Heisenberg spins $\sigma_{k,i} = 1/2$ ($i = 1,2$)
by assuming the periodic boundary conditions.
The interaction constants $I_1$ and $I_2$ label the nearest-neighbor interactions between the nodal Ising spins and
interstitial Heisenberg spins along the sides of the primitive diamond cell,
while the coupling constants $J_1$, $J_2$ and $J_3$ determine the spatially anisotropic \textit{XYZ} interaction between
the nearest-neighbor interstitial Heisenberg spins from the same primitive cell.
Finally, the Zeeman's term $h$ determines the magnetostatic energy of the nodal Ising spins and
interstitial Heisenberg spins in the external magnetic field.
It should be mentioned that a particular case $I_2=0$ (or $I_1=0$) of the Hamiltonian (\ref{htot})
corresponds to the spin-1/2 Ising-Heisenberg doubly decorated chain.

For  further manipulations, it is convenient to rewrite the total Hamiltonian (\ref{htot}) as a sum over the
cell Hamiltonians
\begin{eqnarray}
\hat {\cal H} = \sum_{k=1}^N \hat {\cal H}_k,
\nonumber 
\end{eqnarray}
whereas the cell Hamiltonian $\hat {\cal H}_k$ involves all the interaction terms of the $k$th diamond cell
\begin{eqnarray}
\hat {\cal H}_k &=& J_1 \hat\sigma^x_{k,1} \hat\sigma^x_{k,2} + J_2 \hat\sigma^y_{k,1} \hat\sigma^y_{k,2}
+ J_3 \hat\sigma^z_{k,1} \hat\sigma^z_{k,2}
+ I_1 \left( S_{k}\hat\sigma^z_{k,1} + \hat\sigma^z_{k,2}S_{k+1} \right)\nonumber \\
&+& I_2 \left( S_{k}\hat\sigma^z_{k,2} + \hat\sigma^z_{k,1}S_{k+1} \right)
- h \left(\hat\sigma^z_{k,1} + \hat\sigma^z_{k,2}\right) - \frac{h}{2} \left(S_{k} + S_{k+1}\right).
\nonumber  
\end{eqnarray}
Since the cell Hamiltonians $\hat {\cal H}_k$ commute between themselves,
$\left[\hat {\cal H}_k, \hat {\cal H}_n \right]=0$,
the partition function of the spin-1/2 Ising-Heisenberg diamond chain can be written in such a form:
\begin{eqnarray}
{\cal Z} \equiv \mbox{Tr} \exp \left(-\beta \hat {\cal H} \right)
= \sum_{\{S_k \}} \prod_{k=1}^N {\cal Z}_k(S_{k},S_{k+1}),
\label{pfd}
\end{eqnarray}
where $\beta=1/(k_{\rm B} T)$, and $k_{\rm B}$ is the Boltzmann's constant and $T$ is the absolute temperature,
the symbol $\sum\limits_{\{S_k \}}$ denotes the summation over all possible spin configurations of the nodal Ising spins
and
\begin{eqnarray}
{\cal Z}_k(S_{k},S_{k+1}) &=& \mbox{Tr}_{\{\sigma_{k,1}, ~\sigma_{k,2}\}} \exp \left(-\beta \hat {\cal H}_k \right)
\label{BF}
\end{eqnarray}
is the effective Boltzmann's factor obtained after tracing out the spin degrees of freedom of two Heisenberg spins
from the $k$-th primitive cell.
To proceed further with the calculation, one necessarily needs to evaluate the effective Boltzmann's factor
${\cal Z}_k(S_{k},S_{k+1})$ given by equation (\ref{BF}).
For this purpose, let us pass to the matrix representation of the
cell Hamiltonian $\hat {\cal H}_k$ in the basis spanned over four available states of two Heisenberg spins
$\sigma^z_{k,1}$ and $\sigma^z_{k,2}$:
\begin{eqnarray}
|\!\uparrow, \uparrow \rangle_{k} = \left |\uparrow \right \rangle_{k,1} \left |\uparrow \right \rangle_{k,2},
\quad
|\! \downarrow, \downarrow \rangle_{k} = \left |\downarrow \right \rangle_{k,1} \left |\downarrow \right \rangle_{k,2},
\quad
|\! \uparrow, \downarrow \rangle_{k} = \left |\uparrow \right \rangle_{k,1} \left |\downarrow \right \rangle_{k,2},
\quad
|\!\downarrow, \uparrow \rangle_{k} = \left |\downarrow \right \rangle_{k,1} \left |\uparrow \right \rangle_{k,2},
\label{B}
\end{eqnarray}
whereas $|\!\!\uparrow\rangle_{k,i}$ and $|\!\!\downarrow\rangle_{k,i}$ denote two eigenvectors of the spin operator
$\hat\sigma^z_{k,i}$ with the respective eigenvalues $\sigma^z_{k,i} = 1/2$ and $-1/2$.
After a straightforward diagonalization of the cell Hamiltonian $\hat {\cal H}_k$, one obtains the following four
eigenvalues:
\begin{eqnarray}
{\cal E}_{k\,1,2} (S_{k},S_{k+1}) &=&
\pm \sqrt{\frac{\left(J_1 - J_2\right)^2}{16}  + \left [\frac{I_1 + I_2}{2} \left(S_{k} + S_{k+1}\right) - h\right ]^2}
+ \frac{J_3}{4} - \frac{h}{2} \left(S_{k} + S_{k+1} \right) ,
\nonumber \\
{\cal E}_{k\,3,4} (S_{k},S_{k+1}) &=&
\pm \sqrt{\frac{\left(J_1 + J_2\right)^2}{16}  + \frac{\left(I_1 - I_2\right)^2}{4} \left(S_{k} - S_{k+1}\right)^2}
- \frac{J_3}{4} - \frac{h}{2} \left(S_{k} + S_{k+1} \right) .
\label{Ek}
\end{eqnarray}
Now, one may simply use the eigenvalues (\ref{Ek}) in order to calculate the Boltzmann's factor (\ref{BF}):
\begin{eqnarray}
\hspace{-0.5cm}
{\cal Z}_k(S_{k},S_{k+1})&=&
2\exp \left[\frac{\beta h}{2} \left(S_{k} + S_{k+1} \right)\right]
\nonumber \\
& {\times}&
\left[ \exp \left(\frac{\beta J_3}{4} \right)
\cosh \left(\beta \sqrt{\frac{\left(J_1 {+} J_2\right)^2}{16}  +
\frac{\left(I_1 {-} I_2\right)^2}{4} \left(S_{k} {-} S_{k+1}\right)^2} \right) \right.
\nonumber \\
& {+}& \left.  \exp \left(-\frac{\beta J_3}{4}\right)
\cosh \left(\beta \sqrt{\frac{\left(J_1 {-} J_2\right)^2}{16}  +
\left[\frac{I_1 {+} I_2}{2} \left(S_{k} {+} S_{k+1}\right) {-} h\right]^2} \right)
\right].
\label{bf1}
\end{eqnarray}

The Boltzmann's factor (\ref{bf1}) can be subsequently replaced through the generalized decoration-iteration
transformation \cite{fis59,roj09,str10,bell13} similarly to what was done before \cite{can06,lis3}.
Since this Boltzmann's factor is essentially a transfer matrix [see, equation (\ref{pfd})],
we can directly use the transfer-matrix method \cite{bax82,aps15}.
For convenience we define the elements $V_{ij}$ of the transfer matrix ${\mathbf V}$  as follows:
\begin{eqnarray}
V_{ij} \equiv {\cal Z}_k \left( S_{k}=i-\frac32,~S_{k+1}=j-\frac32 \right).
\label{trm}
\end{eqnarray}
The calculation of the partition function requires finding the eigenvalues of the transfer matrix
\begin{eqnarray}
{\mathbf V} = \left( \begin{array}{ll}
V_{11} & ~V_{12} \\
V_{21} & ~V_{22}
\end{array}
\right).
\nonumber
\end{eqnarray}
The eigenvalues of this matrix are given by the roots of the quadratic characteristic equation.
The expressions for the two transfer-matrix eigenvalues are as follows:
\begin{eqnarray}
\lambda_i = \frac12 \left( V_{11} + V_{22} -
(-1)^i \sqrt{\left(V_{11} - V_{22}\right)^2 + 4V_{12}^2} \right), \quad i=1,2.
\label{VE}
\end{eqnarray}

As a result, the partition function ${\cal Z}$ given by equation (\ref{pfd})
is determined by the transfer-matrix eigenvalues $\lambda_1$ and $\lambda_2$:
\begin{eqnarray}
{\cal Z} = \mbox{Tr} \, {\mathbf V}^N  = \lambda_1^N + \lambda_2^N.
\label{pff}
\end{eqnarray}

Exact results for other thermodynamic quantities follow quite straightforwardly from the formula~(\ref{pff})
for the partition function ${\cal Z}$.
In the thermodynamic limit $N \to \infty$, the free energy per unit cell can be easily obtained:
\begin{eqnarray}
g = -\frac{1}{\beta} \lim_{N \to \infty} \frac{1}{N} \ln {\cal Z} = -\frac{1}{\beta} \ln \lambda_{1},
\nonumber 
\end{eqnarray}
where $\lambda_{1}$ is the largest transfer-matrix eigenvalue.

The entropy $s$ and the specific heat $c$ per unit cell can be subsequently calculated from the formulae
\begin{eqnarray}
s &=& k_{\rm{B}} \beta^2  \frac{\partial g}{\partial \beta}
= k_{\rm{B}} \left( \ln \lambda_{1} - \frac{\beta}{\lambda_{1}} \frac{\partial \lambda_{1}}{\partial \beta}\right) ,
\nonumber\\
c &=& - \beta  \frac{\partial s}{\partial \beta}
= k_{\rm{B}} \beta^2   \left[\frac{1}{\lambda_{1}} \frac{\partial^2 \lambda_{1}}{\partial \beta^2} -
\frac{1}{\lambda_{1}^2} \left(\frac{\partial \lambda_{1}}{\partial \beta}\right)^2 \right],
\nonumber
\end{eqnarray}
whereas the total magnetization $m$ and magnetic susceptibility $\chi$ readily follow from the relations
\begin{eqnarray}
m &=& - \frac{\partial g}{\partial h} = \frac{1}{\beta \lambda_{1}} \frac{\partial \lambda_{1}}{\partial h},
\nonumber \\
\chi &=& \frac{\partial m}{\partial h} =
\frac{1}{\beta} \left[\frac{1}{\lambda_{1}} \frac{\partial^2 \lambda_{1}}{\partial h^2} -
\frac{1}{\lambda_{1}^2} \left(\frac{\partial \lambda_{1}}{\partial h}\right)^2 \right].
\nonumber
\end{eqnarray}
The derivatives $\frac{\partial \lambda_{1}}{\partial \beta}$, $\frac{\partial^2 \lambda_{1}}{\partial \beta^2}$,
$\frac{\partial \lambda_{1}}{\partial h}$, $\frac{\partial^2 \lambda_{1}}{\partial h^2}$
follow straightforwardly from formula (\ref{VE}).

\section{Results and discussions}

Now, let us proceed to the discussion of the most interesting results obtained for
the spin-1/2 Ising-Heisenberg distorted diamond chain with the antiferromagnetic Ising interactions, $I_1>0$ and $I_2>0$,
and ferromagnetic \textit{XXZ} Heisenberg interaction: $J_{3} = J < 0$, $J_{1} = J_{2} = J \Delta$,
where $\Delta$ determines the anisotropy of \textit{XXZ} Heisenberg interaction.
Without loss of generality, we may also assume that one of the two considered Ising couplings is stronger
than the other $I_1 \geqslant I_2$ and introduce the difference between both Ising coupling constants
$\delta = I_1 - I_2\geqslant0$.
To reduce the number of free parameters, we define the following set of dimensionless interaction parameters
\[
\tilde{J}=\frac{J}{I_1},  \qquad \Delta, \qquad \tilde{h}=\frac{h}{I_1},
\qquad \tilde{\delta}=\frac{\delta}{I_1}.
\]
The parameter $\tilde{\delta}$ restricted to the interval $\tilde{\delta} \in [0,1]$
has a physical sense of the distortion parameter, because it determines a relative difference between two
Ising coupling constants $I_1$ and $I_2$.

The ground state of the spin-1/2 Ising-Heisenberg distorted diamond chain can be trivially connected to
the lowest-energy eigenstate of the cell Hamiltonian (\ref{Ek}).
Depending on a mutual competition between the parameters
$\Delta$, $\tilde{J}$, $\tilde{\delta}$ and $\tilde{h}$,
in total one finds four different ground states:
the fully magnetized (FM) state, the ferrimagnetic (FRI) state, the monomer-dimer (MD1) state,
and the quantum antiferromagnetic (QAF1) state given by the eigenvectors
\begin{eqnarray}
|\mbox{FM} \rangle &=& \prod\limits_{k=1}^N| + \rangle_k \otimes |\!\uparrow, \uparrow \rangle_{k},
\nonumber\\
|\mbox{FRI} \rangle &=& \prod\limits_{k=1}^N| - \rangle_k \otimes |\!\uparrow, \uparrow \rangle_{k},
\nonumber\\
|\mbox{MD1} \rangle &=& \prod\limits_{k=1}^N | + \rangle_k \otimes
\frac{1}{\sqrt{2}} \Big(|\!\uparrow, \downarrow \rangle_{k} + |\!\downarrow, \uparrow \rangle_{k} \Big),
\nonumber\\
|\mbox{QAF1} \rangle &=& \left\{\begin{array}{l}
\prod\limits_{k=1}^N \left|[-]^k \right\rangle_k \otimes
\left(A_{[-]^{k+1}} |\!\uparrow, \downarrow \rangle_{k} + A_{[-]^k} |\!\downarrow, \uparrow \rangle_{k} \right)
\\
\prod\limits_{k=1}^N \left|[-]^{k+1} \right\rangle_k \otimes
\left(A_{[-]^k} |\!\uparrow, \downarrow \rangle_{k} + A_{[-]^{k+1}} |\!\downarrow, \uparrow \rangle_{k} \right)
\end{array} \right. .
\label{GS}
\end{eqnarray}
In the above, the ket vector $|\pm\rangle_k$ determines the state of the nodal Ising spin $S_k = \pm 1/2$,
the symbol $[-]^k \in \{-,+\}$ denotes the sign of the number $(-1)^k$,
the spin states relevant to two Heisenberg spins from the $k$th primitive cell are defined by the notation (\ref{B}),
and the probability amplitudes $A_{\pm}$ are explicitly given by the expressions
\begin{eqnarray}
A_{\pm}=\frac{1}{\sqrt{2}}
\sqrt{1 \pm \frac{\tilde{\delta}}{\sqrt{\left(\tilde{J} \Delta\right)^2 + \tilde{\delta}^2}}}.
\nonumber
\end{eqnarray}
The eigenenergies per unit cell that correspond to the respective ground states (\ref{GS}) are given as follows:
\begin{eqnarray}
\tilde{\cal E}_{\rm{FM}} &=& \frac{\tilde{J}}{4} + 1 - \frac{\tilde{\delta}}{2} - \frac{3\tilde{h}}{2},
\nonumber\\
\tilde{\cal E}_{\rm{FRI}} &=& \frac{\tilde{J}}{4} - 1 + \frac{\tilde{\delta}}{2} - \frac{\tilde{h}}{2},
\nonumber\\
\tilde{\cal E}_{\rm{MD1}} &=& -\frac{\tilde{J}}{4} + \frac{1}{2}\tilde{J}\Delta  - \frac{\tilde{h}}{2},
\nonumber\\
\tilde{\cal E}_{\rm{QAF1}} &=& -\frac{\tilde{J}}{4} -
\frac{1}{2}\sqrt{\left(\tilde{J} \Delta \right)^2 + \tilde{\delta}^2}.
\label{EE}
\end{eqnarray}

\begin{figure}[!t]
\begin{center}
\includegraphics[width=0.95\textwidth]{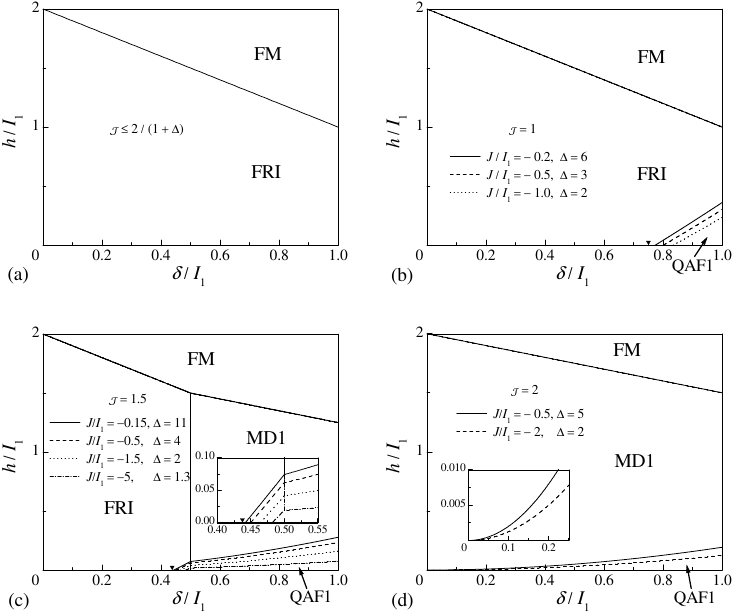}
\end{center}
\caption{The ground-state phase diagram in the $\tilde{\delta} - \tilde{h}$ plane for a few typical values of
the ferromagnetic Heisenberg coupling constant ($\tilde{J}<0$, $\Delta$),
which are consistent with four different topologies depending on the value of the parameters $\cal{J}$ and $\Delta$:
(a) ${\cal{J}} \leqslant 2/(1+\Delta)$; (b) ${\cal{J}} = 1.0$; (c) ${\cal{J}} = 1.5$; (d) ${\cal{J}} = 2.0$.
The symbols $\blacktriangledown$ shown in figure~\ref{fig2}b and figure~\ref{fig2}c allocate limiting
$\Delta \to \infty$ position of the point, at which the ground states FRI and QAF1 coexist at zero field.}
\label{fig2}
\end{figure}

The ground-state phase diagram in the $\tilde{\delta}-\tilde{h}$ plane can have four different topologies depending
on the parameters ${\cal{J}} = |\tilde{J}|(\Delta - 1)$ and $\Delta$ (see, figure~\ref{fig2}).
It should be noted for further analysis that the parameter ${\cal{J}}$ must agree with the parameter $\Delta$,
namely, the value $\Delta = 1$ corresponds to the value ${\cal{J}} = 0$,
the condition $\Delta<1$ ($\Delta>1$) corresponds to the condition ${\cal{J}}<0$ (${\cal{J}}>0$).
The first type of the ground-state phase diagram (figure~\ref{fig2}a) is found for ${\cal{J}} \leqslant 2/(1+\Delta)$.
According to this condition, in the case of $0\leqslant\Delta\leqslant1$,
only the first type of the ground-state phase diagram is realized.
It includes only two ground states FM and FRI.
The second type of the ground-state phase diagram (figure~\ref{fig2}b) is realized for
$2/(1+\Delta)<{\cal{J}} \leqslant 1$ and it involves three ground states: FM, FRI, and QAF1.
In the zero field, the FRI and QAF1 states are separated by the point
\[
\tilde{\delta} = \tilde{\delta}_{\rm{F}\cdot\rm{Q1}} \equiv \frac{2-{\cal{J}}}{2}
\left(1 + \frac{{\cal{J}} \Delta}{2(\Delta - 1) + {\cal{J}}} \right).
\]
The third type of the ground-state phase diagram (figure~\ref{fig2}c) involving all four available ground states
FM, FRI, QAF1, and MD1 can be detected for $1< {\cal{J}} <2$.
The FRI and MD1 ground states are separated by the phase boundary
$ \tilde{\delta} = \tilde{\delta}_{\rm{F}|\rm{M1}} \equiv 2 - {\cal{J}} $.
It is easy to show that $\tilde{\delta}_{\rm{F}\cdot\rm{Q1}} \leqslant \tilde{\delta}_{\rm{F}|\rm{M1}}$.
In the regime of fixedly ${\cal{J}}$  the parameter $\tilde{\delta}_{\rm{F}\cdot\rm{Q1}}$ is restricted
to the interval $\left(1 - {\cal{J}}^2/4, 2-{\cal{J}}\right)$
as the exchange anisotropy varies from $\Delta \rightarrow \infty$ to $\Delta \rightarrow 1$.
The fourth type of the ground-state phase diagram (figure~\ref{fig2}d) emerges for ${\cal{J}} \geqslant 2$
and it includes three ground states: FM, MD1, and QAF1.
The phase boundary between the QAF1 and MD1 states starts from the special point $(0,0)$,
which corresponds to the highly frustrated (FRU1) ground state
\[
|\mbox{FRU1} \rangle = \prod\limits_{k=1}^N
\Big(\left|+\right\rangle_k \mbox{or} \left|-\right\rangle_k \Big) \otimes
\frac{1}{\sqrt{2}} \Big(|\uparrow, \downarrow \rangle_{k} + |\downarrow, \uparrow \rangle_{k} \Big),
\]
for ${\cal{J}} > 2$.
And if ${\cal{J}} = 2$, then the two ground states FRU1 and FRI coexist together at the point $(0,0)$.
The FRU1 ground state has the residual entropy $s_{\rm{res}} = k_{\rm B}\ln 2$ reflecting
the macroscopic degeneracy $2^N$, which comes from spin degrees of freedom of the nodal Ising spins.

The topologies of the ground-state phase diagram b, c, d (see, figure~\ref{fig2})
are identical to the topologies of the phase diagrams of the ground state of the first, second, and third type
of the corresponding antiferromagnetic chain in \cite{lis3}.
At the same time, the ground states have some differences:
in the MD1 state, the spins on the Heisenberg bond are in the triplet state,
and in the corresponding state of the antiferromagnetic chain, they are in the singlet state \cite{lis3}.
In the QAF1 state, the spins on the Heisenberg bond are in a symmetric superposition
of quasiclassical antiferromagnetic states,
and in the corresponding state of the antiferromagnetic chain, they are in an antisymmetric superposition of
quasiclassical antiferromagnetic states \cite{lis3}.
We see that under the condition $2/(1+\Delta)<{\cal{J}}$, the properties of the ground state
of our antiferromagnetic-ferromagnetic chain without geometric frustration are analogous
to the properties of the ground state of the corresponding antiferromagnetic chain with geometric frustration.
Based on this, it can be concluded that in this chain,
under the condition $2/(1+\Delta)<{\cal{J}}$, an effective geometric frustration appears.
Under the condition ${\cal{J}}>1$, the topology of the ground-state phase diagrams in the plane
$\tilde{\delta}-\tilde{h}$ (phase diagrams c and d in figure~\ref{fig2}) defines only the ${\cal{J}}$ parameter,
which under this condition has the meaning of the topology parameter.
The topology parameter ${\cal{J}}$ determines the type of phase diagram of our chain similarly to
the topology parameter $|\tilde{J}|(\Delta + 1)$ in the corresponding antiferromagnetic chain \cite{lis3}.
This means that under the condition ${\cal{J}}>1$, the effective geometric frustration
most fully reproduces the manifestations of the usual geometric frustration in the properties
of ground state of the spin-1/2 Ising-Heisenberg distorted diamond chain.
The effective geometric frustration in our antiferromagnetic-ferromagnetic chain arises due to quantum fluctuations.
It happens as follows: if the \textit{XX}-interaction (quantum fluctuations) is stronger than the \textit{ZZ}-interaction
on the Heisenberg ferromagnetic \textit{XXZ} bond, the pair of spins of this bond has the lowest energy in the quantum state,
which is a symmetric superposition of quasi-classical antiferromagnetic states
$|\! \uparrow, \downarrow \rangle$ and $|\! \downarrow, \uparrow \rangle$ according to (\ref{B}).

\begin{figure}[!t]
\begin{center}
\includegraphics[width=0.5\textwidth]{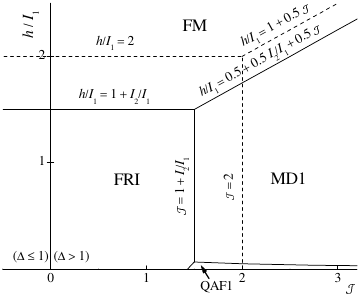}
\end{center}
\caption{The ground-state phase diagram in the ${\cal{J}} - \tilde{h}$ plane,
general schematic version, which is obtained at $\Delta=5$ and $\tilde{I}_2=0.5$.
The mathematical formulae for the ground-state phase boundaries are explicitly quoted along the relevant lines.
In addition to this, the phase boundary between the FRI and QAF1 ground states is
$\tilde{h}=\sqrt{{\cal{J}}^2 \Delta^2/(\Delta-1)^2+(1-\tilde{I}_2)^2} - {\cal{J}}/(\Delta-1) - 1 - \tilde{I}_2$,
and the phase boundary between the MD1 and QAF1 ground states is
$\tilde{h}=\sqrt{{\cal{J}}^2 \Delta^2/(\Delta-1)^2+(1-\tilde{I}_2)^2} - {\cal{J}}\Delta/(\Delta-1)$.
The dashed lines correspond to the symmetric case of the Ising interaction $\tilde{I}_2=1$,
in which the QAF1 ground state is absent.}
\label{fig3}
\end{figure}

The general form of the ground-state phase diagram can be obtained in the ${\cal{J}}-\tilde{h}$ plane (figure~\ref{fig3}).
Under this circumstance, the boundary-phase point between the FRI and QAF1 ground states at zero field is given by
\begin{eqnarray}
{\cal{J}}_{\rm{F}\cdot\rm{Q1}} =
\frac{1}{\Delta+1} \left(1 + \tilde{I}_2 + \sqrt{\left( 1-\tilde{I}_2 \right)^2 + 4 \tilde{I}_2 \Delta^2} \right),
\nonumber
\end{eqnarray}
where $\tilde{I}_2 = I_2/I_1$.
It should be noticed that the FRU1 ground state does exist in the relevant ground-state phase diagram along the line
given by $\tilde{h}=0$ and ${\cal{J}}>2$ if the complete symmetry $\tilde{I}_2=1$ is recovered due to the absence
of the QAF1 ground state.

Next, let us examine the thermodynamic characteristics as a function of the temperature, the magnetic fields,
and the distortion parameter. To illustrate all possible scenarios,
we have selected the values of the Heisenberg coupling constants $\tilde{J}$ and $\Delta$ in regime ${\cal J} =$~const
in order to fall into the parameter region pertinent to the ground-state phase diagrams shown in figure~\ref{fig2}c
involving all available ground states.

The behavior of the magnetization depending on the field for different temperatures and
on the temperature for different fields is identical to
the behavior of the magnetization of the corresponding antiferromagnetic chain,
which was discussed in detail in~\cite{lis3}.
In particular, the field dependence of magnetization $m/m_{\rm{s}}$ at zero temperature,
where $m_{\rm{s}}=3/2$ is the saturation magnetization,
can have a zero plateau $m/m_{\rm{s}}=0$ corresponding to the QAF1 ground state,
an intermediate plateau $m/m_ {\rm{s}}=1/3$ corresponding to the FRI ground state or the MD1 ground state,
and the saturation plateau $m/m_{\rm{s}}=1$ corresponding to the FM ground state.
The temperature curve of magnetization in the region of medium and high temperatures shifts upwards
with the strengthening of quantum fluctuations.

The magnetic susceptibility multiplied by the temperature ($\chi k_{\textrm{B}} T$)
as a function of the temperature in the zero field
also behaves similarly to the corresponding antiferromagnetic chain \cite{lis3}.
In particular, as the temperature tends to zero, the quantity $\chi k_{\textrm{B}} T$ can either
exponentially diverge as in quantum ferrimagnetics \cite{yam99},
if $\tilde{\delta}$ corresponds to the FRI ground state, or exponentially tends to zero,
if $\tilde{\delta}$ corresponds to the QAF1 ground state, or takes the value of 1/4,
if $\tilde{\delta}$ is at the point of coexistence of the states FRI and QAF1.
The high-temperature dependence of $\chi k_{\textrm{B}} T$ shifts to a higher susceptibility
with the strengthening of quantum fluctuations.

\begin{figure}[!tb]
\begin{center}
\includegraphics[width=0.85\textwidth]{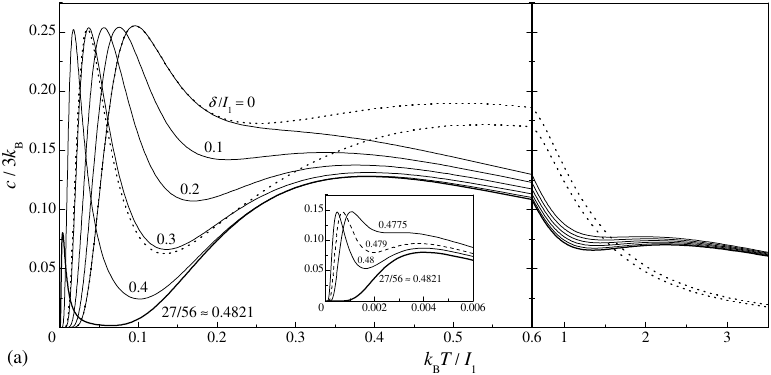} \\
\includegraphics[width=0.85\textwidth]{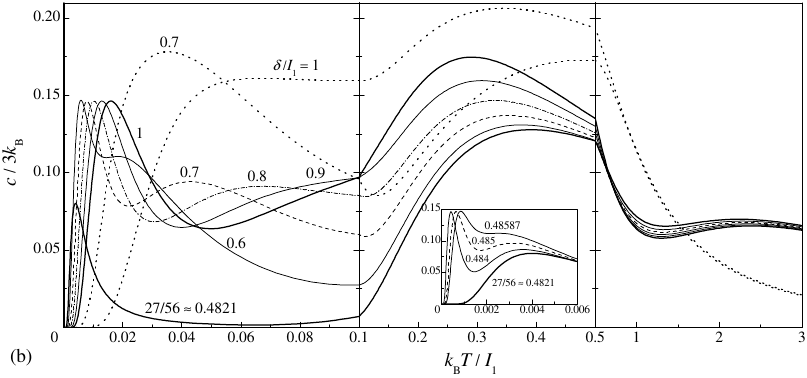}
\end{center}
\caption{The temperature dependencies of the zero-field specific heat for the ferromagnetic Heisenberg interaction
$\tilde{J}=-5$, $\Delta=1.3$ (${\cal{J}}=1.5$) and several values of the distortion parameter $\tilde{\delta}$:
(a) $\tilde{\delta}\leqslant \tilde{\delta}_{\rm{F}\cdot\rm{Q1}} = 27/56 \approx 0.4821$;
(b) $\tilde{\delta}\geqslant \tilde{\delta}_{\rm{F}\cdot\rm{Q1}}$.
The dotted lines correspond to another particular case $\tilde{J}=-0.5$, $\Delta=4$ (${\cal{J}}=1.5$),
which serve in evidence of the role of quantum fluctuations in the ${\cal{J}}$-constant regime.
The insets show, in an enlargened scale, two low-temperature peaks emerging close to the phase boundary between the
FRI and QAF1 ground states. }
\label{fig4}
\end{figure}

Let us consider the temperature dependence of the zero-field specific heat
in the case of a strong Heisenberg ferromagnetic interaction $\tilde{J}=-5$ and $\Delta=1.3$ (${\cal{J}}=1.5$),
which is presented in figure~\ref{fig4} for different values of the distortion parameter $\tilde{\delta}$.
In order to understand a variation of the temperature dependence of the zero-field specific heat
when the parameters of the Heisenberg ferromagnetic interaction are changed,
we also show the results for the case,
when the Heisenberg interaction in the same regime (${\cal{J}}=1.5$) is much weaker,
namely $\tilde{J}=-0.5$, $\Delta=4$ (figure~\ref{fig4}).
As we can see in figure~\ref{fig4}, the heat capacity temperature curve has the main round maximum,
which is located in the region of average temperatures (around $k_{\rm{B}} T / I_1 = 0.4 $),
and can have one or two additional low-temperature peaks.
This profile of the heat capacity temperature curve in the zero field is analogous
to the typical profile of the heat capacity temperature curve in the corresponding antiferromagnetic chain \cite{lis3}.
Note that when the Heisenberg interaction is strong, then at very small distortions $\tilde{\delta} \cong 0$,
the heat capacity can have only one maximum (the main and additional maxima merge)
for the entire range of low and medium temperatures [see the curve for $\tilde{\delta} = 0$ in figure~\ref{fig4}(a)].
An interesting feature of the heat capacity temperature curve in the case of a strong Heisenberg interaction
is an additional broad maximum in the region of very high temperatures (somewhat above $k_{\rm{B}} T / I_1 = 2 $),
which exists for the entire range of values of the distortion parameter $\tilde{\delta}$ (figure~\ref{fig4}).
The height of this maximum slightly decreases due to the growth of the distortion parameter $\tilde{\delta}$.
Analysis of the structure of the energy spectrum of the cell Hamiltonian (\ref{Ek}) showed
that the strong Heisenberg ferromagnetic interaction significantly distances
the group of nearby highest energies ${\cal E}_{k\,3} (\pm,\pm)$ from the rest of the energies.
Based on this, we can conclude that the high-temperature additional maximum is formed by the high-energy excitations
of cell spins to the states with the energies ${\cal E}_{k\,3}(\pm,\pm)$.
It should be noted that in the temperature dependence of the zero-field specific heat
of the corresponding antiferromagnetic chain, no additional high-temperature maximum was found \cite{lis3},
and the temperature curve of the heat capacity of the Ising-Hubbard distorted diamond chain
in the case of geometric frustration has a similar additional maximum \cite{lis11-1}.

Let us consider more in detail the transformation of the temperature dependent heat capacity
under the influence of the distortion parameter $\tilde{\delta}$, which varies in the range of possible values $[0,1]$.
If the distortion parameter increases in the region $[0,\tilde{\delta}_{\rm{F}\cdot\rm{Q1}})$,
then the main maximum of the heat capacity loses its height and shifts to higher temperatures,
and  the additional low-temperature peak shifts to lower temperatures.
When the parameter $\tilde{\delta}$ falls into a certain small neighborhood
of the point $\tilde{\delta}_{\rm{F}\cdot\rm{Q1}}$,
then the additional low-temperature peak splits into two maxima [see the inset in figure~\ref{fig4}(a)].
Upon further convergence of the parameter $\tilde{\delta}$ with the point $\tilde{\delta}_{\rm{F}\cdot\rm{Q1}}$,
the left-hand low-temperature peak moves to zero temperature and disappears at
$\tilde {\delta}=\tilde{\delta}_{\rm{F}\cdot\rm{Q1}}$.
When the parameter $\tilde{\delta}$ is above the point $\tilde{\delta}_{\rm{F}\cdot\rm{Q1}}$,
the left-hand low-temperature peak appears again near the zero temperature.
As the distortion increases, the left-hand low-temperature peak moves toward higher temperatures and
merges with the right-hand low-temperature maximum [see inset on the figure~\ref{fig4}(b)].
Therefore, the left-hand low-temperature peak is formed by thermal excitations between the FRI and QAF1 states.
And the right-hand low-temperature peak is formed by thermal excitations between the FRI and MD1 states and
between the QAF1 and MD1 states.
With further growth of the $\tilde{\delta}$ parameter, the low-temperature peak moves toward higher temperatures,
and the main maximum increases its height and moves toward lower temperatures.
It should also be noted that the growing distortion in the region $(\tilde{\delta}_{\rm{F}\cdot\rm{Q1}},1]$ shifts
the heat capacity curve upwards on the temperature segment near the main maximum, but at higher temperatures,
this effect is opposite, that is in the region of high temperatures,
the growing distortion parameter shifts the heat capacity curve down.

In the case of a strong Heisenberg ferromagnetic interaction given by the parameters $\tilde{J}=-5$ and $\Delta=1.3$,
we discovered another splitting of the low-temperature peak of the zero-field specific heat [see figure~\ref{fig4}(b)],
which occurs when the parameter $\tilde{\delta}$ changes in the range of values (0.6, 0.84),
which is far from the point $\tilde{\delta}_{\rm{F}\cdot\rm{Q1}}$.
After the splitting of the low-temperature peak, as the distortion increases,
the right-hand low-temperature maximum approaches the main maximum and merges with it at $\tilde{\delta}=0.84$.
By analyzing the dependence of the energies of the QAF1, FRI, and MD1 states on the distortion parameter and
Heisenberg interaction parameters, it can be established that the left-hand low-temperature peak is formed by
thermal excitations between the QAF1 and MD1 states, and the right-hand low-temperature maximum is formed by
thermal excitations between the QAF1 and FRI states and between the MD1 and FRI states.
It should be noted that this splitting of the low-temperature  peak of the heat capacity occurs with
a strong Heisenberg interaction, which ensures the necessary proximity of the energies of the QAF1 and MD1 states,
and a rather strong distortion, which ensures the necessary distance between the energy of the FRI state and
the energy of the ground state QAF1. A weak Heisenberg interaction, for example $\tilde{J}=-0.5$, $\Delta=4$ (${\cal{J}}=1.5$),
does not give such a splitting of the low-temperature maximum [figure~\ref{fig4}(b)].

\begin{figure}[!tb]
\begin{center}
\includegraphics[width=0.95\textwidth]{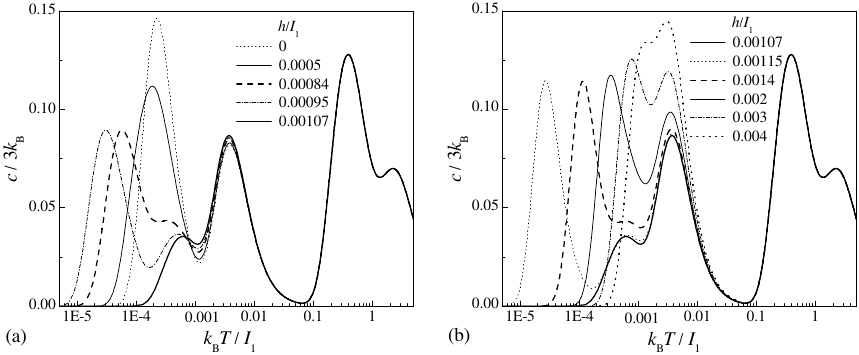}
\end{center}
\caption{The semi-logarithmic plot of temperature dependencies of the specific heat
for the ferromagnetic Heisenberg interaction $\tilde{J}=-5$, $\Delta=1.3$ (${\cal{J}}=1.5$),
the distortion parameter $\tilde{\delta} = \tilde{\delta}_{\rm{F}\cdot\rm{Q1}} + 0.001 \approx 0.4831$,
and some selected values of the magnetic field $\tilde{h}$ within the interval $[0,0.004]$:
(a) $\tilde{h} \in [0, 0.00107]$ and (b) $\tilde{h} \in [0.00107, 0.004]$.
}
\label{fig5}
\end{figure}

\begin{figure}[!tb]
\begin{center}
\includegraphics[width=0.5\textwidth]{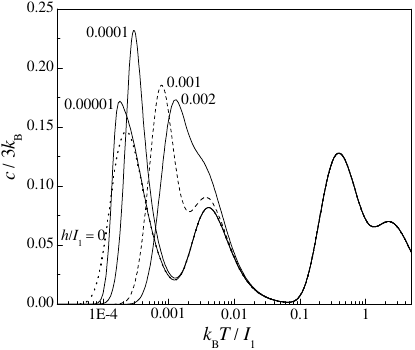}
\end{center}
\caption{The semi-logarithmic plot of temperature dependencies of the specific heat
for the ferromagnetic Heisenberg interaction $\tilde{J}=-5$, $\Delta=1.3$ (${\cal{J}}=1.5$),
the distortion parameter
$\tilde{\delta}=\tilde{\delta}_{\rm{F}\cdot\rm{Q1}} - 0.001 \approx 0.4811$,
and some selected values of the magnetic field $\tilde{h}$ within interval $[0,0.002]$.
}
\label{fig6}
\end{figure}

Last but not least, let us examine the temperature variations of the specific heat in the presence of non-zero
external magnetic field.
In particular, we study the effect of a weak magnetic field on the temperature dependence of the heat capacity
for a distortion very close to the point $\tilde{\delta}_{\rm{F}\cdot\rm{Q1}}$,
at which the temperature curve of the zero-field heat capacity has four maxima.
For this purpose, we consider the effect of a weak magnetic field on the temperature dependence of heat capacity
in two cases:
in the first case we start from the point $(0,\tilde{\delta}_{\rm{F}\cdot\rm{Q1}} {+} 0.001)$
in the region of the ground state QAF1 [figure~\ref{fig2}(c)],
and in the second case we start from the point $(0,\tilde{\delta}_{\rm{F}\cdot\rm{Q1}} {-} 0.001)$
in the region of the FRI ground state [figure~\ref{fig2}(c)].
In the first case (see figure~\ref{fig5}), the additional low-temperature heat capacity maximum,
which is formed by thermal excitations between the FRI and QAF1 states,
splits when the magnetic field increases [figure~\ref{fig5}(a)] and
the temperature dependence of the heat capacity has five maxima.
After splitting, the left-hand peak shifts to lower temperatures and disappears when the value of the magnetic field
corresponds to the point on the line of coexistence of the QAF1 and FRI states [figure~\ref{fig2}(c)].
As soon as the point $(\tilde{h},\tilde{\delta}_{\rm{F}\cdot\rm{Q1}} {+} 0.001)$
moves into the FRI state region [figure~\ref{fig2}(c)],
an additional peak occurs near zero temperature [figure~\ref{fig5}(b)].
With further strengthening of the magnetic field, this peak shifts toward higher temperatures and
merges with the neighboring maximum [figure~\ref{fig5}(b)].
The maximum formed as a result of the merger continues to shift toward higher temperatures as the magnetic field increases
and merges with the neighboring maximum at $\tilde{h}\geqslant0.004$ [figure~\ref{fig5}(b)].
In the second case (figure~\ref{fig6}), the additional low-temperature heat capacity maximum does not split
in a weak magnetic field. It only deforms and shifts towards higher temperatures when the magnetic field increases.
At the value of the magnetic field $\tilde{h}=0.002$, the additional low-temperature heat capacity maximum merges
with the neighboring maximum (figure~\ref{fig6}).
In order to understand the mechanism of influence of a weak magnetic 
field on the temperature dependence of
heat capacity in the case when the distortion is very close to the point $\tilde{\delta}_{\rm{F}\cdot\rm{Q1}}$,
consider the influence of a weak magnetic field on the energy of the FRI state
(the energy of the QAF1 state does not depend on the magnetic field).
In a zero magnetic field, the ferrimagnetic state FRI has the same energy as the ferrimagnetic state
$|\mbox{FRI}1\rangle = \prod\limits_{k=1}^N| + \rangle_k \otimes |\!\downarrow, \downarrow \rangle_{k}$,
and in a weak magnetic field, the energies of the FRI and FRI1 states are different.
As a result of splitting, the energy of the ferrimagnetic states FRI and FRI1 by a very weak magnetic field,
the low-temperature maximum, which in a zero magnetic field is formed by thermal excitations
between the states FRI and QAF1, can split under certain conditions.
In the first case (figure~\ref{fig5}), the low-temperature maximum splits since a very weak magnetic field changes the QAF1 ground state to the FRI ground state [figure~\ref{fig2}(c)].
In the second case (figure~\ref{fig6}), the low-temperature maximum does not split,
because a very weak magnetic field does not change the FRI ground state [figure~\ref{fig2}(c)].
\newpage

\section{Conclusions}

We have rigorously examined the ground state and thermodynamics
of the spin-1/2 Ising-Heisenberg distorted diamond chain within the transfer-matrix method.
Our attention was focused on how the distortion and ferromagnetic \textit{XXZ} Heisenberg interaction
affects the magnetization process and specific heat of the spin-1/2 Ising-Heisenberg distorted diamond chain
with the antiferromagnetic Ising interactions and the ferromagnetic \textit{XXZ} Heisenberg interaction.
In this case, there is no geometric spin frustration in the given chain.
We show that quantum fluctuations on a Heisenberg ferromagnetic bond create an effective geometrical spin frustration
when a pair of spins of this bond is in an energetically favorable quantum antiferromagnetic state.
Then, the ground state and thermodynamic characteristics of our antiferromagnetic-ferromagnetic chain have the
features similar to the corresponding antiferromagnetic chain with geometric spin frustration \cite{lis3}.

The ground-state phase diagram of the spin-1/2 Ising-Heisenberg distorted diamond chain
with the antiferromagnetic Ising interactions and ferromagnetic Heisenberg interaction
totally consists of four different ground states:
the fully magnetized state FM, the ferrimagnetic state FRI, the monomer-dimer state MD1,
and the quantum antiferromagnetic state QAF1.
The quantum ground states MD1 and QAF1 emerging for the ferromagnetic Heisenberg interaction differ
from the analogous quantum ground states emerging for the antiferromagnetic Heisenberg interaction just by
antisymmetric and symmetric quantum superposition of two antiferromagnetic states of the Heisenberg spin pairs,
respectively.
The ground-state phase diagram in the distortion parameter --- magnetic field plane
can have four different topologies depending on the ferromagnetic Heisenberg coupling parameters.
The second, third and fourth topologies of this ground-state phase diagram,
which are realized under the condition $2/(1+\Delta)<{\cal{J}}$,
are the same as the three topologies of the ground-state phase diagram of the antiferromagnetic chain~\cite{lis3}.
The third and fourth topologies of the ground-state phase diagram,
which are realized under the condition $1<{\cal{J}}$,
are defined by the topology parameter $\cal{J}$, which is analogous to
the topology parameter in the antiferromagnetic chain \cite{lis3}.
Therefore, under the condition $2/(1+\Delta)<{\cal{J}}$, an effective geometric frustration occurs
in our antiferromagnetic-ferromagnetic chain,
which is most fully manifested in the properties of the ground state under the condition $1<\cal{J}$.

The effective geometrical spin frustration affects the magnetization, magnetic susceptibility,
and heat capacity of an antiferromagnetic-ferromagnetic chain similarly to the geometrical spin frustration
in an antiferromagnetic chain \cite{lis3}.
In particular, the magnetization curve of the spin-1/2 Ising-Heisenberg distorted diamond chain
may involve at most two different intermediate plateaus at zero and 1/3 of the saturation magnetization.
The distortion parameter is responsible for a rich variety of temperature dependencies
of the specific heat, which may display one or two low-temperature peaks in addition
to the main round maximum observable at medium temperatures.
The strong Heisenberg interaction is responsible for an additional broad maximum in the temperature dependence
of heat capacity in the region of very high temperatures.
The combination of strong Heisenberg interaction and significant distortion can split
the peak in the low-temperature heat capacity.
If the distortion parameter is quite close to the right of the point $\tilde{\delta}_{\rm{F}\cdot\rm{Q1}}$,
then a very weak magnetic field can split the low-temperature maximum of the heat capacity and
then the temperature dependence of the heat capacity has five maxima.
The physical origin of all observed low-temperature and hight-temperature peaks of the specific-heat has been clarified
on the grounds of relevant thermal excitations.

It is worthwhile to remark that the investigated spin system reduces to the spin-1/2 Ising-Heisenberg
doubly decorated chain in the particular case $I_2=0$ ($\tilde{\delta}=1$)
and the symmetric spin-1/2 Ising-Heisenberg diamond chain in the other particular case $I_1=I_2$
($\tilde{\delta}=0$).

\section*{Acknowledgements}

The author is grateful to T.~Verkholyak for the discussion and useful remarks.

{\small \topsep 0.6ex

}

\newpage
\ukrainianpart

\title{Спін-1/2 ромбічноподібний ланцюжок Iзiнга-Гайзенберга з
антиферомагнітними взаємодіями Iзiнга і феромагнітною взаємодією Гайзенберга}
\author{Б. М. Лісний}
\address{Інститут фізики конденсованих систем Національної академії наук України,
вул. Свєнціцького, 1, 79011, Львів, Україна}
%
%
%

\makeukrtitle

\begin{abstract}
\tolerance=3000%
Точно розв'язуваний спін-1/2 ромбічноподібний ланцюжок Iзiнга-Гайзенберга у зовнішньому магнітному полі
досліджено у випадку антиферомагнітної взаємодії Ізінга і феромагнітної \textit{XXZ} взаємодії Гайзенберга.
Детально вивчено вплив квантових флуктуацій і дисторсії (асиметрії) на основний стан,
магнітні та теплові властивості моделі.
Зокрема, встановлено, що крива намагнічування при нульовій температурі може мати
проміжні плато при нульовій намагніченості та при 1/3 намагніченості насичення.
Показано, що температурна залежність питомої теплоємності може мати чотири максимуми при нульовому магнітному полі
та п'ять максимумів при слабкому магнітному полі.
Фізичне походження всіх спостережуваних додаткових піків питомої теплоємності було з'ясовано на основі
домінуючих теплових збуджень.
Ми показали, що квантові флуктуації породжують ефективну геометричну фрустрацію в цьому ланцюжку.
\keywords ромбiчноподiбний ланцюжок Iзiнга-Гайзенберга, основний стан, фазова діаграма, питома теплоємність

\end{abstract}

\end{document}